# Core Graduate Courses: A Missed Learning Opportunity?

Chandralekha Singh and Alexandru Maries

*Department of Physics and Astronomy, University of Pittsburgh, Pittsburgh, PA 15260*

**Abstract.** An important goal of graduate physics core courses is to help students develop expertise in problem solving and improve their reasoning and meta-cognitive skills. We explore the conceptual difficulties of physics graduate students by administering conceptual problems on topics covered in undergraduate physics courses before and after instruction in related first year core graduate courses. Here, we focus on physics graduate students' difficulties manifested by their performance on two qualitative problems involving diagrammatic representation of vector fields. Some graduate students had great difficulty in recognizing whether the diagrams of the vector fields had divergence and/or curl but they had no difficulty computing the divergence and curl of the vector fields mathematically. We also conducted individual discussions with various faculty members who regularly teach first year graduate physics core courses about the goals of these courses and the performance of graduate students on the conceptual problems after related instruction in core courses.

**PACs:** 01.40.Fk, 01.40.Ha

## INTRODUCTION

An important goal of a graduate core physics course is to enable students to develop complex reasoning and problem solving skills and use these skills in a unified manner to explain and predict diverse phenomena [1]. In physics, there are few fundamental laws. They are expressed in compact mathematical forms and can provide students tools for organizing their knowledge hierarchically. Good organization of knowledge is crucial for easy retention and retrieval of knowledge and can help students in reasoning and deciding which concept is applicable in a particular context. Conceptual reasoning and sense making of principles defined precisely in a quantitative form is at the heart of developing a robust knowledge structure and becoming an expert in physics. A crucial difference between the problem solving strategies used by experts in physics and those who are learning to become experts lies in the interplay between how the knowledge is organized and how it is retrieved to solve problems. Experts are comfortable going between different knowledge representations, e.g., mathematical, diagrammatic, etc., and employ representations that make problem solving easier [2].

In order to build a robust knowledge structure of physics using complex mathematical tools, graduate students must not treat quantitative problem solving in core courses merely as a mathematical exercise but as an opportunity for sense making and they must learn to draw conceptual inferences from quantitative tools. Unfortunately, conceptual reasoning which is important for developing expertise in physics is under-emphasized in traditional physics courses all the way from the introductory level to the graduate level. A majority of students at all levels have competing priorities for time and they are not necessarily intrinsically motivated to learn (although one may expect graduate students to be more motivated than introductory physics students many of whom are non-majors) and focus on learning what they are assessed on. Therefore, a lack of emphasis on conceptual assessment in physics courses may be an obstacle in the road towards expertise and hinder the development of a robust knowledge structure and a functional understanding.

A survey at the First Conference on Graduate Education in Physics co-organized by one of the co-authors suggests that unlike other science and engineering departments, a majority of physics departments with a Ph.D. program currently require that all graduate students take physics core courses and perform adequately in those courses [3]. An important goal of the core graduate courses is to help students expand upon the knowledge acquired earlier in the undergraduate courses to develop a deeper functional understanding of the fundamental physics principles. An implicit assumption in requiring that the graduate students perform better than a set cut-off level in these core courses (and/or a comprehensive exam based on the material covered in the core courses) to continue in the Ph.D. program is that students will acquire functional knowledge in these courses that they would be able to use in their future research.

It is noteworthy that for many decades a majority of faculty members teaching these core courses have used the same textbooks (e.g., the book by Jackson for electricity and magnetism, the book by Goldstein for classical mechanics) nationwide [3]. A discussion with the faculty members teaching these courses at the University of Pittsburgh (Pitt) suggests that they felt that these textbooks have the core of what the students need to learn and at the level they need it for their future research in different subfields of physics. A review of the final exams that some of the faculty members at Pitt had administered in the last few years suggests that there is great uniformity in what they expected graduate students to be able to do on the exams. Regardless of the faculty member, all of the final exam questions were quantitative, requiring complex calculations.

## GOAL AND METHODOLOGY

Our research goal was to explore the conceptual difficulties of physics graduate students by administering qualitative problems on topics covered in a typical upper-level undergraduate physics course before and after instruction in related first year core graduate courses.

The conceptual questions, a majority in the multiple-choice format, some of which required students to explain their reasoning for their choices were administered to first year physics graduate students enrolled in core courses at the beginning of instruction as part of pre-tests used to determine whether the students should remain in the graduate core courses or take the corresponding undergraduate course first. The pre-tests administered to graduate students had other problems synthesized by the instructors of those courses. The conceptual problems given as a post-test were administered after related instruction in a core course as a part of an exam later in the course. Here, we discuss only two conceptual questions related to divergence and curl of vector fields in diagrammatic representation which were administered in a graduate core electricity and magnetism (EM) course. We note that these questions related to fields discussed below are not specifically about electric or magnetic fields but, since they are related to vector fields, EM was deemed to be the most appropriate course in which to administer them (similar figures can be found in the undergraduate EM textbook by Griffiths which was used at Pitt). Also, administering the "post-test" after related instruction does not imply that the instructor discussed these types of conceptual questions explicitly in those courses. We call it "post-test" only in that we hypothesize that the students may review these concepts on their own and/or be reminded of these concepts while preparing for the more mathematically challenging homework and exams and attending lectures in the related core courses.

Apart from the written conceptual problems administered to graduate students mostly in the multiple-choice format (out of which only two representative problems will be discussed here), we discussed the problems individually with a subset of graduate students. We also discussed their opinions about what they learned in the graduate core courses and what those courses were able to achieve. Although seven faculty at Pitt responded to a written survey about core courses, better insight was obtained when we discussed individually with nine faculty members their views about the goals of the core courses and the performance of graduate students on the conceptual problems administered after related topics were covered in the core courses. Although the discussions were not recorded, we took notes immediately after each discussion to capture the thought processes.

Here, we focus on graduate students' performance on two representative qualitative problems involving diagrammatic representation of vector fields as follows:

1. The figure below shows a cross-section of a field in the y-z plane. The field has no dependence on the x coordinate. Longer arrows represent stronger field.

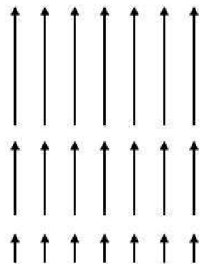

(a) The above field has a non-zero divergence but zero curl.
(b) The above field has a zero divergence but non-zero curl.
(c) The above field has both non-zero divergence and non-zero curl.
(d) The above field has both zero divergence and zero curl.

2. The figure below shows a cross-section of a field in the x-y plane. The field has no dependence on the z coordinate. Longer arrows represent stronger field.

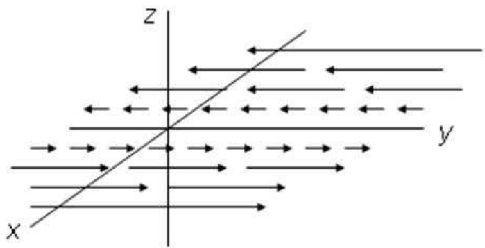

(a) The above field has a non-zero divergence but zero curl.
(b) The above field has a zero divergence but non-zero curl.
(c) The above field has both non-zero divergence and non-zero curl.
(d) The above field has both zero divergence and zero curl.
(e) Impossible to predict.

## RESULTS

Table 1 shows physics graduate students' performance on these problems before and after instruction in the EM core course. It shows that at the beginning (pre) of the EM graduate core course, on each question, approximately half of the graduate students answered correctly and at the end of the course (post), two-thirds of the graduate students answered correctly. The poor performance of graduate students before the core course may partly be due to the fact that there was a significant time interval between when they were administered the pre-test and when they took the undergraduate course in which these concepts were discussed. It is encouraging that their performance improved at the end of the EM core course. However, on each question, approximately one third of the students answered the question incorrectly in the post-test. Table 1 shows that the most common distracter in both Problems 1 and 2 for graduate students before instruction in the core course was that the field in each diagram has both zero divergence and zero curl (option (d)). In individual discussions, some students explained that the confusion related to the first diagram (Problem 1) stemmed from the fact that field lines were not spreading out from a point source. One way to correctly reason about this diagram is to note that if one chooses a closed volume, weaker field lines go in and stronger field lines come out denoting a non-zero divergence. For students who selected option (d) for Problem 2, the confusion regarding zero curl often originated from the fact that they expected that a non-zero curl would imply that the field lines would be curved (e.g., in a circle or an ellipse) or that the field lines have an inversion symmetry about the y-z plane implying no divergence or curl of the field.

**TABLE 1.** The percentage of graduate students who chose options (a)-(e) on the multiple choice questions discussed. The correct response for each question is italicized. The numbers in parentheses in the second column refer to the total number of students in that group.

| Q | Group | A | b | c | d | e |
|---|---|---|---|---|---|---|
| 1 | Grad (Pre) (37) | *49* | 0 | 11 | 35 | 5 |
|   | Grad (Post) (18) | *67* | 11 | 11 | 11 | 0 |
| 2 | Grad (Pre) (37) | 3 | *46* | 16 | 27 | 8 |
|   | Grad (Post) (18) | 11 | *61* | 17 | 11 | 0 |

When we individually asked some of the students to calculate the divergence and curl of some vector fields, we found that all of them were able to calculate the divergence and curl correctly without difficulty. In other words, they had difficulty in recognizing if the diagrams of the vector field have non-zero divergence and curl but they had no difficulty in calculating the divergence and curl of a field which can be computed in an algorithmic manner without understanding conceptually what a non-zero divergence or curl of a field may look like. This difficulty in dealing with the diagrammatic representation of divergence and curl in Problems 1 and 2 suggests that graduate students have conceptual difficulties with divergence and curl even though they know how to calculate them mathematically for a given vector field.

We then discussed with individual faculty members the goals of the physics graduate core courses and the graduate students' performance on the conceptual questions before and after instruction in related core courses. A few instructors noted that they spend time in mathematical sense making and making conceptual connections between the mathematics and physics concepts as relevant while others noted that graduate students should make the time for such connections themselves due to time-constraints in the course. However, individual discussions with physics graduate students in those courses suggests that they did not feel that there was much focus on concepts in any of the graduate core courses which focused on making sure that they were able to solve mathematically challenging problems assigned in homework and examinations. Some of the graduate students specifically noted that the focus of the core courses was exclusively quantitative and they felt bogged down with the mathematical manipulations introduced at a tremendous pace in all of their core courses. Therefore, they did not process the material conceptually.

One graduate course instructor admitted that he did not expect that students will improve significantly from pre-test to post-test after the core courses on conceptual topics because his course focused on helping students learn to solve mathematically rigorous problems. We (the re-searchers) expressed concern that since new knowledge builds on prior knowledge, if graduate students did not get an adequate opportunity to connect what they were learning in the graduate core courses with the concepts they learned in their undergraduate courses, they may not be acquiring functional knowledge that can be retrieved after the core courses are over and they pursue their research. The faculty member replied that graduate students should come prepared with concepts covered in the undergraduate courses because they are prerequisite for the graduate core courses and students in core courses should take the time to connect the mathematics and physics on their own as needed. He insisted that the graduate core courses cannot focus on concepts because they must teach students how to solve mathematically challenging problems beyond what is taught at the undergraduate level.

Individual discussions with core course instructors also suggests that the final exam problems were generally like the homework problems but often easier in that they required shorter calculations than what is required to solve a typical homework problem on those topics so that students could finish the exam in a few hours (typically 2-4 hours were assigned by the Pitt instructors for the final exam in core courses). Some of the instructors specifically admitted that many of the homework and exam questions they used in the core course they taught were the same or similar to what they were asked to do when they were graduate students many years ago. It appears from the discussions with the core course instructors that they felt that if the core courses have prepared them well for their career as a physics professor, they should also prepare their students well. In designing these core courses, the instructors did not worry much about the increased diversity in the student population in the graduate physics programs over the years, the fact that many students will choose non-academic careers and the physics graduate programs are becoming more interdisciplinary over the years with varying needs of the students. A majority of the instructors also felt that there is pressure to cover a lot of material in a short amount of time but that if they covered a certain material, a majority of students should be able to develop a functional understanding of that material by paying attention to their lectures. Most instructors noted that very few graduate students came to ask for help during office hours.

When we asked instructors if they would be willing to assign graduate students integrated conceptual and quantitative problems (problems in which students solve quantitative problems similar to those they were already assigned but include some conceptual questions that help students to contemplate qualitative implications of the quantitative procedures involved), some faculty members were more enthusiastic than others about including such problems in homework. All of them expressed concern about the fact that the textbooks they currently use in the core courses do not include such problems and they do not have the time to construct such problems. Some of them expressed concern about the fact that their homework assignments are already long and including conceptual questions may increase the time taken to finish the assignment and increase the workload graduate students have. We argued that assigning integrated conceptual and quantitative problems as homework in graduate core courses will not increase the time to complete the assignment significantly but some of them were still skeptical about whether they would include conceptual questions on homework and exams if they were available or whether graduate students should take the time to do such mathematical sense-making on their own. It appeared that these instructors did not necessarily discern value in assigning such problems. However, most instructors were willing to assign such problems in homework if good problems were made available to them (although they admitted that they did not have the time to make such questions). Some of them were also willing to assess graduate students in core courses using both conceptual and quantitative problems if students had worked on both types of problems in homework.

## DISCUSSION AND SUMMARY

Although the performance of graduate students on the problems related to divergence and curl discussed here improved after the EM core course, on each question, approximately one third of the students were unable to answer correctly. Individual discussions with instructors of core courses suggest that their major goal is to teach students how to solve more mathematically challenging problems than those learned in undergraduate courses. They expected that students should make conceptual connections on their own but did not assess their conceptual understanding. In our previous research related to graduate students' attitudes and approaches to problem solving, one survey question asked whether equations must be intuitive in order to use them or whether they routinely use equations even if they are non-intuitive [4]. In response to this question, only slightly more than 50% of the graduate students noted that when answering graduate level problems, equations must be understood in an intuitive sense before being used. The following sample response from a graduate student reflects their sentiments: "I am often still presented with equations to calculate something without enough motivation to understand the process, even at the graduate level, and being able to use the equation and accept that you'll understand it later is often necessary..."

We argue that physics graduate core courses in their current format may be a missed learning opportunity because they focus exclusively on making graduate students mathematically facile but ignore whether they are developing a functional understanding of the underlying concepts adequately. As a community, we must consider the goals of physics graduate core courses and also contemplate the assessment tools that will inform whether we have met those goals.

## ACKNOWLEDGMENTS

We thank the National Science Foundation for support.